# Regulating Large Language Models

## A Roundtable Report

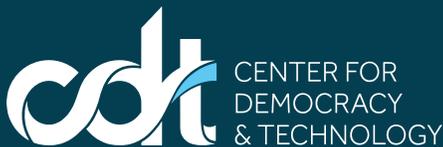

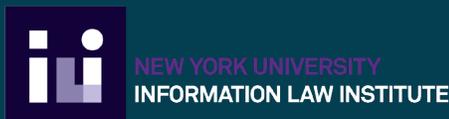

**Paul Friedl**
**Gabriel Nicholas**

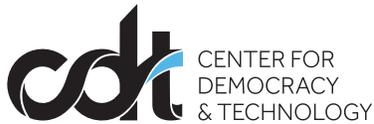

The **Center for Democracy & Technology (CDT)** is the leading nonpartisan, nonprofit organization fighting to advance civil rights and civil liberties in the digital age. We shape technology policy, governance, and design with a focus on equity and democratic values. Established in 1994, CDT has been a trusted advocate for digital rights since the earliest days of the internet. The organization is headquartered in Washington, D.C. and has a Europe Office in Brussels, Belgium.

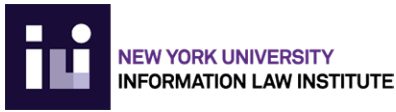

The **NYU Information Law Institute** is an academic center for the study of law, policy, and social norms defining and affecting the flow of information in a digitally networked society. Its mission is to encourage and disseminate thoughtful research and commentary, welcoming the participation of faculty, students, and other researchers across the disciplinary spectrum.

# Regulating Large Language Models

## A Roundtable Report

**Paul Friedl and Gabriel Nicholas**

With contributions from Miranda Bogen, Samir Jain, Katherine Strandburg, and Dhanaraj Thakur.

*Acknowledgements*

We would like to thank Nicole Artz for her invaluable assistance in organizing this workshop. We also thank all of the attendees for their valuable time and comments. We have left out their names and affiliations here because this event was conducted under Chatham House Rules. Thank you to Tim Hoagland for copy-editing and design.

This work was made possible with support from the NYU Information Law Institute and through a grant from the John S. and James L. Knight Foundation.

*Suggested Citation*

Friedl, P. & Nicholas, G. (2024). "Regulating Large Language Models: A Roundtable Report." New York University Information Law Institute and Center for Democracy & Technology. https://cdt.org/insights/regulating-large-language-models-a-roundtable-report/

# Contents

# 01 Introduction

On July 20, 2023, a group of 27 scholars and digital rights advocates with expertise in law, computer science, political science, and other disciplines gathered for the Large Language Models, Law, and Policy Roundtable, co-hosted by the **NYU School of Law Information Law Institute** and the **Center for Democracy & Technology**. The roundtable convened to discuss how law and policy can help address some of the larger societal problems posed by large language models (LLMs). The discussion focused on three policy topic areas in particular:

- *Truthfulness*: What risks do LLMs pose in terms of generating mis- and disinformation? How can these risks be mitigated from a technical and/or regulatory perspective?

- *Privacy*: What are the biggest privacy risks involved in the creation, deployment, and use of LLMs? How can these risks be mitigated from a technical and/or regulatory perspective?

- *Market concentration*: What threats do LLMs pose concerning market/power concentration? How can these risks be mitigated from a technical and/or regulatory perspective?

For each area, we first had two experts — one from computer science, the other from law or policy — introduce the topic and highlight what they identified to be the most pressing challenges. We then held an open, fishbowl-style conversation about the question posed. The day concluded with a brainstorming session asking participants to debate the potentials and downsides of different regulatory interventions that could be used to address the problems raised earlier in the day.

**Co-hosted by the NYU School of Law Information Law Institute and the Center for Democracy & Technology**

In this paper, we provide a detailed summary of the day's proceedings. We first recap what we deem to be the most important contributions made during the issue framing discussions. We then provide a list of potential legal and regulatory interventions generated during the brainstorming discussions. The Roundtable was run under Chatham House rules so this paper will neither disclose the names of participants nor feature direct quotes or attributions.

# 02

# Policy Topic Areas

## Truthfulness

In this first session, discussion leaders reviewed the definitional, technical, and sociological challenges of getting language models to generate more truthful information. Our non-technical discussion leader provided a **taxonomy of types of untruthful text**, distinguishing between misinformation (unintentionally incorrect text), disinformation (text created to convince people of something false), propaganda (text created to convince people of an idea), and biased information (text that reflects prejudice contained in the training data). Expounding on the first category, our technical discussion leader explained three factors that can lead language models to generate misinformation: incorrect training data, outdated training data, or "hallucinating", i.e. when a model generates realistic sounding but untrue statements because they are not grounded in truth, but rather in statistical linguistic patterns.[1]

Discussion leaders also highlighted potential strategies for mitigating untruthfulness in language models. These strategies varied according to the type and cause of untruthfulness. For

---

instance, preventing language models from reproducing incorrect or out of date information in their training data could be achieved by directing models toward more trusted sources of information, e.g. through better prompting, finetuning, or retrieval augmentation.[2] Interventions to mitigate propaganda, on the other hand, may not affect the model itself, but rather, focus on undermining propaganda agents' abilities to disseminate content and affect peoples' beliefs, such as through social media.[3] Other interventions discussion leaders mentioned included spreading radioactive data to make generated content more detectable, building digital provenance standards, creating usage limitations, and investing in public media literacy. All interventions, however, would need to grapple with the difficult issue of finding **shared criteria for what is "truthful,"** a topic that is simultaneously complex, subjective, and political.

Group discussion of this topic began with a debate over whether the problems language models introduce to the information environment are novel. One participant recalled Wikipedia and Google causing similar "misinformation panics" that never came to fruition. Once people become more literate in how language models work and better understand their limits, the participant argued, LLMs will become less disruptive to the larger information ecosystem. Other participants however highlighted certain qualities of LLMs that could **undermine the effectiveness of technical literacy alone**. First, people may not always know when they are interacting with a language model. Second, today's language models rarely if ever provide users with the necessary sources or citations to enable them to validate or contextualize the information these models produce. Third, LLMs' limitations may vary widely across contexts. Even if a user has general knowledge of language models' limitations, they may not know of the specific shortcomings within the specific context in which they are interacting with it. As LLMs take on a wider range of roles (e.g. HR personnel, medical professional, content moderator), this challenge of understanding their limitations will only increase.

---

2   See e.g "What is retrieval-augmented generation?" *IBM Blog*, 22 August 2023. [perma.cc/MRU9-CG87]

3   Goldstein, Josh A., et al. "Generative language models and automated influence operations: Emerging threats and potential mitigations." *arXiv preprint arXiv:2301.04246* (2023). [perma.cc/S34H-R3U7]

At the same time, participants raised concerns that policymakers could **over-index on LLM-related misinformation concerns**. In many ways, one participant argued, private interests align with improving models' truthfulness, particularly to enable potentially lucrative enterprise software use cases. Participants in general expressed interest in policymakers exploring more sectoral and specific risks of untruthful language model outputs rather than imposing broad "truthfulness obligations." Interventions that would dictate certain standards or sources by which to identify and correct "misinformation" would likely face significant challenges under the U.S.' First Amendment, and other global free speech protections.

# Privacy

Our legal discussion leader first explained that there are at least three ways language models can violate individuals' privacy: by training on private data, by outputting private data, or by aggregating previously dispersed data and thereby deriving new, "emergent" private information."[4] Our technical discussion leader then focused on the second of these methods, speaking about the privacy concerns that arise from large language models' capacity for reproducing their training data verbatim, a phenomenon known as **memorization**.[5] Due to memorization, models may end up leaking personally identifiable information (PII), thereby compromising individuals' privacy. Our discussion leader then described three different technical approaches to mitigating memorization-based privacy concerns, each with its own tradeoffs and limitations:

1. Developers can try to **remove private information from training data sets**. This, however, presupposes that developers are capable of consistently identifying personally identifiable/private data, which is rarely the case since there is no one format or content of private data. Removing private information could also degrade the performance of the model in unexpected ways. For instance, if developers were to delete all zip codes from a model's training data, this could also undermine valuable non-PII-related use cases.

---

4    Participants disagreed on whether Large Language Models were indeed able to create derivative or "emergent" information.

5    Biderman, Stella, et al. "Emergent and predictable memorization in large language models." *arXiv preprint arXiv:2304.11158* (2023) [perma.cc/LAN7-6AG8]; recent research has also shown that memorization becomes worse the bigger a model is, see Carlini, Nicholas, et al. "Quantifying memorization across neural language models." *arXiv preprint arXiv:2202.07646* (2022) [perma.cc/7DZE-PKMX].

2. Developers can implement **differential privacy**, an approach whereby controlled noise is added to the training data to protect individuals' privacy while maintaining certain statistical axioms to assure the data is still useful and accurate.[6] Differential privacy however, too, is predicated on developers' ability to consistently identify private information and relies on a number of formal assumptions that rarely hold true for large unstructured text databases.[7]

3. Developers can apply **output filtering**, i.e. explicitly forbid models from outputting certain types of data or content. In addition to the above-mentioned difficulties in identifying PII, output filters have been shown to be vulnerable to exploits and malicious work-arounds.

Whereas all these technical approaches to alleviating memorization-based privacy harms exhibit certain technical limitations, there is also a more fundamental issue complicating the discussion: it is **unclear which information should be considered private** in the first place. Both our discussion leaders suggested that the **contextual integrity theory of privacy** could help answer these challenging questions regarding what and when data should be considered private in LLM development and deployment. Contextual integrity asserts that information flows impinge upon privacy when they violate the information norms governing a specific context,[8] i.e. where they are contextually inappropriate. The theory hence does not relieve policymakers from exploring what LLM-related information uses communities deem appropriate.

---

6   For a recent literature review, see Hu, Lijie, et al. "Differentially Private Natural Language Models: Recent Advances and Future Directions." *arXiv preprint arXiv:2301.09112* (2023). [perma.cc/4ACN-7BDK]

7   See also Brown, Hannah, et al. "What does it mean for a language model to preserve privacy?" *Proceedings of the 2022 ACM Conference on Fairness, Accountability, and Transparency.* 2022. [perma.cc/M9CZ-BBM6]

8   See Nissenbaum, Helen. *Privacy in context: Technology, policy, and the integrity of social life.* Stanford University Press, 2020. [perma.cc/WWE9-P7VL]

It does, however, establish a generative theoretical framework, introducing important conceptual considerations[9] that can inform productive analyses and provoke comparison with privacy norms established in (potentially) similar contexts such as search engines. For instance, contextual integrity's idea of "transmission principles" raises questions such as, do LLM developers need to ensure individuals' consent when processing personal data? What tools or protocols could help them do so?[10] Or can specific methods of presenting personal information, such as including links to source websites, make it more appropriate?

The discussion started with participants asking why LLM developers could not simply delete or replace easily identifiable PII, such as personal names or phone numbers. Other participants retorted that such an approach could have unexpected knock-on effects, including systems having difficulties representing the trajectories of individuals over time. In general, participants highlighted **potential trade-offs between privacy and conventional metrics of performance**.

Discussants also debated **whether privacy regulations should pivot away from their focus on PII in favor of a more risk- or harms-based approach**. Use of PII would then only be regulated/prohibited where such use could result in material harm to individuals (e.g. higher prices or discriminatory treatment in consumer credit applications). Others, however, dismissed these proposals, arguing that privacy was important regardless of potential downstream harms and that people have a right to privacy even when it is not "quantifiable." Still others warned that privacy concerns could lead to chilling effects, as internet users may withdraw to non-public fora out of fear over the mining of their personal communications.

---

9    Contextual integrity asserts that privacy norms can be described with reference to five conceptual parameters: 1) Information subject (Who is the information about?), 2) Sender (From whom does the information originate?), 3) Recipient (Who receives the information?), 4) Information type (What type of information is transmitted or used?), and 5) Transmission principle (Under what circumstances may information be transmitted or used? *Ibid*.

10   For more, see *infra* Section 3, "Regulating web-scraping."

Participants suggested two possible regulatory interventions that they believed should be explored in greater depth. First, some argued that **reporting mechanisms** could allow users to flag privacy-violating outputs, putting LLM providers in a position to prevent these infringements. Second, participants advocated for **intra-industry knowledge sharing structures**, through which LLM-providers could inform each other of attack vectors or vulnerabilities they have identified.[11] Multiple participants agreed that market mechanisms alone have generally failed to satisfy consumers' demand for privacy, and will likely continue to do so in the realm of LLMs without strong, clearly delineated regulatory interventions. As noted by one participant, new institutional structures, such as citizen councils, may be needed to enable a quicker democratic development of privacy norms for new technologies.[12]

> **As powerful chips are still in short supply, a large share of today's most efficient chips are held by a very small number of large companies.**

# Market Concentration

Our technical discussion leader began by offering a framework of the **LLM application supply chain divided into (at least) five layers**:[13] 1) a data layer, comprised of companies constructing and managing training datasets at scale (e.g. Common Crawl); 2) a compute layer, comprised of companies offering computational resources (e.g. AWS); 3) a foundation model layer, comprised of companies training LLMs (e.g. OpenAI); 4) a hosting layer, comprised of companies hosting LLMs (e.g. Hugging Face); and 5) an application layer, comprised of companies building user applications on top of existing LLMs (e.g. chatbot services). Any LLM-based application thus relies on resources and inputs from at least five different markets. Concentration can occur in any one of them.

---

11    For more, see *infra* Section 3, "Intra-industry information sharing frameworks."

12    For a recent policy brief on public consultation processes for AI development, see Gilman, Michele. "Democratizing AI: Principles for Meaningful Public Participation." *Data & Society*, 27 September 2023. [perma.cc/7S3D-FSFP]

13    This model of the LLM supply chain was introduced in Jones, Elliot. "Explainer: What is a foundation model?" *Ada Lovelace Institute,* 17 July 2023. [perma.cc/DRF6-STMY]

Our technical discussion leader then discussed one potential harm that could stem from market concentration: **algorithmic monocultures**. An algorithmic monoculture exists when multiple decision-makers use different algorithmic systems that leverage the same foundation model. Individuals subject to decisions made by different systems can then end up experiencing the same, potentially negative and biased outcome. For instance, job applicants applying to multiple jobs may encounter the same biases across different hiring algorithms, leading to the systematic discrimination of certain candidates.[14]

Our legal discussion leader first argued that meaningful discussion of AI policy issues always requires acknowledging the effects of market concentration. For instance, the high demand of compute power, driven primarily by the current scarcity of chips, means that cloud **computing providers have great power over the development of AI technologies**. Regulators including the FTC[15] and Ofcom[16] are carefully monitoring these dynamics in ongoing antitrust investigations.

Finally, our discussion leader emphasized a **mutually-reinforcing relationship between market concentration and privacy harms**. Companies turn to privacy-infringing tactics in order to gain a data advantage over their competitors. Once they have achieved a dominant market position, they are then able to leverage that position to engage in even greater data extraction without needing to fear user exodus or other negative consequences.

The open discussion started with attendees exchanging views on the concentration of compute power. As powerful chips are still in short supply, a **large share of today's most efficient chips** (above all NVIDIA's A100 and H100 GPUs) **are held by a very small number of large companies** (i.e. Amazon, Google, Meta, xAI, and

---

Microsoft). Attendees mentioned that companies were already trying to leverage this advantage to lock customers into cloud infrastructures by bundling access to compute with other services and products. Other attendees, however, noted that the cost of building a midsize language model was still relatively low and accessible for many companies.

Discussion then turned to the second basic resource behind LLMs: data. Participants expressed concern that **market dominance in the LLM space could be self-reinforcing**, as popular services would be able to retrain their models on data gathered from user interactions. Other participants, however, believed that often, user data was less valuable than imagined. What was more important was having the financial means to buy non-public data from data enrichment services, such as Surge AI, Scale, and Appen. Participants pointed to training data services as another area of concentration, arguing that these three companies dominated the markets for data labeling, annotation, and curation. One participant suggested that such market concentration could further worsen conditions for "data laborers," whose work already is precarious and exploitative.

> **Participants also expressed concern that a highly concentrated market for LLMs could create too few points of failure.**

Multiple participants expressed concern that **AI regulation could exacerbate market concentration** and suspected that existing incumbents are trying to raise regulatory costs as a tactic to solidify their market position and impose barriers to entry. Participants also expressed concern that a highly concentrated market for LLMs could create too few points of failure. Another participant argued that monopolization could also lead to less calculable harms, such as undermining diversity and democracy writ large. For these reasons, some participants argued that regulatory proposals must include measures to actively counteract monopolization.

Participants also hotly debated the competitive effects of "open source" models, in particular Meta's LLaMA suite. Participants argued that **many of these "open source" models may not be sufficiently accessible for competitors to make practical use of.** Limitations include 1) licenses that often restrict commercial uses, especially uses that could threaten the developer's own products and interests; 2) limited information about the models training code and data, and 3) storage and compute demands that are so high that they practically forbid others from making use of the model, and compute demands for those who seek to run state of the art models. Participants largely agreed that for many of these models, these limitations make the term "open source" misleading.

# 03 Regulatory Interventions

In the last part of our workshop, participants were asked to ideate and discuss potential regulatory interventions to address the discussed topic areas. This raised a number of questions regarding the most effective regulatory approaches to LLM governance, presented here.

Most participants agreed that one central challenge lies in the fact that developers, deployers, and regulators have **insufficient information to tackle** some of the above-mentioned problems effectively. Large language models are complex systems that can exhibit peculiar, often unpredictable behavior. Which prompts will trigger discriminatory or privacy-compromising outputs, or which vulnerabilities malicious actors might be able to exploit can often only be discovered by interacting with the system. Participants therefore agreed that regulators should think about interventions that could effectively help raise the vulnerability knowledge of LLM providers.

Participants suggested that **user complaint mechanisms**, such as those introduced by the EU's platform-focused Digital Services Act, could improve LLM developers' understanding of the problems their products may raise. Notice mechanisms — which some LLM providers have already begun to implement — can help detect important shortcomings, such as discriminatory outputs or clear misinformation. It is less clear, however, how providers should need to deal with incoming complaints, since notice mechanisms might be exploited and not every question has an objective answer (e.g. when considering issues of truthfulness). One approach participants suggested could lie in obliging LLM-providers to implement a "reasonable complaint-handling system." LLM-providers would then not need to prove that decisions or actions have been taken regarding every single complaint, but would need to prove that they have implemented suitable mechanisms to respond to the most critical or most wide-spread complaints. Responsible vulnerability disclosure policies, such as those established for cybersecurity vulnerabilities, and bug bounty programs for algorithmic harms could help pressure providers into fixing bugs while simultaneously avoiding collateral damage created through the premature publication of unpatched vulnerabilities.[17]

> **Most participants agreed that one central challenge lies in the fact that developers, deployers, and regulators have insufficient information to tackle some of the [thorniest] problems effectively.**

Participants also discussed how **red-teaming** offers another, more established mechanism for surfacing risks associated with LLMs. Most major LLM-providers have already instituted red teaming processes in one way or another, which help facilitate the discovery of both security vulnerabilities as well as harmful or undesired outputs.[18] Attendees explored ways in which regulatory

---

17  For a detailed recent report on bug bounties for algorithmic harms, see Kenway, Josh et al., "Bug Bounties for Algorithmic Harms." 2022. Algorithmic Justice League, January, 2022 [perma.cc/E4PD-UC2W]; Shen, Hong, et al. "Everyday algorithm auditing: Understanding the power of everyday users in surfacing harmful algorithmic behaviors." *Proceedings of the ACM on Human-Computer Interaction* 5.CSCW2 (2021): 1-29. [perma.cc/2S59-VMWJ]

18  See e.g., Ganguli, Deep, et al. "Red teaming language models to reduce harms: Methods, scaling behaviors, and lessons learned." *arXiv preprint arXiv:2209.07858* (2022) (Anthropic). [perma.cc/8WNS-9XP5] interventions could enhance the effectiveness of red-teaming. Suggestions included establishing minimum funding requirements, ensuring the institutional independence of red-teamers from other teams in the company, mandating that recommendations arising from red-teaming activities are binding on company executives, and stipulating that companies must share red-teaming insights with oversight bodies. While many of the duties proposed by the EU's AI Act would practically demand the institution of a red team,[19] some of these suggestions have perhaps not been considered sufficiently.

**Intra-industry information sharing frameworks** were also discussed as an option to help establish effective communication pathways through which LLM-providers could benefit from the internal monitoring activities of other providers. Participants mentioned several examples of industry collaboration efforts that could provide inspiration, such as the anti-CSAM network clustered around the National Center for Missing & Exploited Children's (NCMEC) hash databases or the anti-disinformation network organized by the EU Code of Practice on Disinformation. Information-sharing networks could also lead to establishing best practices and standardizing red-teaming. Other participants, however, flagged potential risks, such as the creation of disincentives for providers to red-team aggressively,[20] or dangers related to the sharing of vulnerabilities that have not been patched fully.

Participants also debated the role of **transparency and transparency obligations**. Some argued that transparency could increase the risk awareness of both LLM providers and third parties, such as oversight bodies, independent researchers, downstream deployers, and end users, helping them better understand the societal risks these models pose. Others warned of the danger of "transparency washing," and highlighted that many approaches to transparency could fail to yield meaningful, actionable insights. Participants generally agreed that any transparency obligations would have to be tailored to their intended goals and audience. Many participants expressed that they saw a strong need for more transparency about companies' training data. They argued that information regarding the sources of training data could enable researchers to have a better understanding of the potential shortcomings of a specific system, and could also reduce barriers for bringing (legitimate) legal actions.

---

19    See e.g. Article 28b(2) of the European Parliament's proposal. European Parliament, *Artificial Intelligence Act*, A9-0188/2023 (2023). [perma.cc/68TM-V3J8]

20    The idea here is that providers will refrain from aggressive red-teaming if they a) have to fear public criticism for their findings or b) are unable to gain a competitive advantage over other providers if they can equally benefit from the results.

**Any transparency obligations would have to be tailored to their intended goals and audience.**

Other discussed interventions did not so much focus on generating actionable information, but rather attempted to mitigate certain policy problems directly.

**Watermarking artificially generated content** was one such discussed strategy.[21] Watermarking could help to make AI-generated content detectable even after distribution, thereby potentially enabling other actors (e.g. social media platforms or even end-users) to identify misinformation or automated influence campaigns. However, although most dominant LLM companies recently pledged to develop "robust watermarking mechanisms,"[22] most roundtable participants agreed that reliable watermarking techniques may still be far off, especially for text.

Finally, some participants discussed whether **regulating web-scraping** could help address privacy and copyright concerns. Indeed, currently pending copyright litigation[23] might end up

---

limiting the legality of scraping copyrighted content.[24] Restricting web-scraping may restrict LLM-developers' currently unfettered access to web-hosted personal information. However, it could also have collateral damage, hampering research, reducing models' capability, and threatening small businesses that use web-scraping to compete with large incumbents. Participants suggested that it could be worth exploring new technical and legal mechanisms for individuals to give consent on whether and when "their" data may be used for training LLMs. Such mechanisms could include updated robot exclusion protocols[25] or new licensing and compensation structures.[26]

# 04 Conclusion

The aim of the Large Language Models, Law, & Policy Roundtable was to explore the threats language models pose to society and ideate about how law and policy could be used to address them. As expected, the event did not end with consensus on a fully fleshed out regulatory paradigm. However, in addition to the topics and regulatory proposals discussed above, the roundtable also yielded a number of important general takeaways on how to approach and frame regulatory conversations about large language models.

First, we found that **comparisons to earlier technologies and their regulatory regimes** helped shed light on which concerns about large language models are novel and which are not. Participants compared language models to search engines, databases, social media platforms, and other technologies. While no comparison was one-to-one, the process of questioning them helped prioritize the most pressing issues and identify those that require more research to address. The same applied for existing regulatory approaches. For instance, discussion of why torts law has so far largely failed to protect individuals from privacy harms elucidated why it might also fail to address the individual harms of language models.

Second, the **internal complexity of language models makes it difficult to know what existing laws should apply to them and how.** Even when researchers have full access to training data and model weights, it is nearly impossible to determine with certainty why a model produces a given output.[27] This creates challenges for assigning legal responsibilities, as it can be unclear whether unexpected or undesirable outcomes are due to "flawed" training data, model construction, finetuning, or usage. This, for instance, complicates the application of liability law, and free speech, and CDA Section 230, which rely on concepts such as "communicative intent," "knowledge," and "reckless disregard."[28] These problems will only become more difficult as language models grow more technically complex and if companies reduce model access.[29] Participants thus also agreed that, despite all difficulties, scientific communities, and computer science in particular, should put a stronger focus on more "fundamental" research into the inner workings of LLMs, investigating causalities and more structural mitigation strategies.

Third, language models may be new, but they are not green fields of regulation. Participants frequently mentioned how U.S. policymakers have treated language models as a chance to remedy the mistakes made in the regulation (or lack of regulation) regarding social media by creating laws before dynamics become too entrenched. This perspective however ignores that **powerful economic and sociotechnical dynamics have already taken hold.** Language models are used by millions every day, have been incorporated into countless businesses, and are in the process of reshaping many jobs. Development pipelines, stretching from the scraping of large

---

volumes of data to the institution of finetuning mechanisms, have emerged and largely consolidated. The competitive landscape too has already somewhat solidified as a few powerful actors hold the lion's share of the best chips and have already reserved much of future production. Finally, publicly available, "open source" language models already allow new market entrants, but are also being used to harmful ends, such as generating mis/disinformation. Regulation that fails to account for these existing dynamics risks failing its ends.

Perhaps most importantly, the Large Language Models, Law, and Policy Roundtable itself highlighted the **importance and effectiveness of interdisciplinary collaboration**. LLMs touch many aspects of society, and neither technologists, social scientists, nor law and policy experts are equipped to mitigate their potential harms on their own. Figuring out what issues to address, where in the stack to intervene, and how to account for individual context will always require a multitude of expertises. We hope that the Roundtable can serve as a model for future initiatives, so as to better harness the collective knowledge of experts to address these important emerging issues.

cdt.org

cdt.org/contact

**Center for Democracy & Technology**
1401 K Street NW, Suite 200
Washington, D.C. 20005

202-637-9800

@CenDemTech